\documentstyle[prl,aps,psfig]{revtex}

\draft

\def\ga{\mathrel{\mathpalette\fun >}}
\def\fun#1#2{\lower3.6pt\vbox{\baselineskip0pt\lineskip.9pt
  \ialign{$\mathsurround=0pt#1\hfil##\hfil$\crcr#2\crcr\sim\crcr}}}
\newcommand{\beq}{\begin{equation}}
\newcommand{\eeq}{\end{equation}}
\begin{document}
\twocolumn[\hsize\textwidth\columnwidth\hsize\csname
@twocolumnfalse\endcsname


\draft
\title{Comment on ``On the Origin of the Highest Energy Cosmic Rays''}

\author{L. N. Epele and E. Roulet}

\address{Departamento de F\'\i sica, Universidad Nacional de La 
Plata, CC67 (1900) La Plata, Argentina}


\maketitle

\vskip 2pc]


In the paper ``On the origin of the highest energy cosmic rays'' 
\cite{st97}, 
Stecker has analysed the implications of new estimates of the 
extragalactic infrared (IR) background \cite{ma98} 
on the attenuation of the 
ultra--high energy cosmic ray (UHCR) nuclei, trying in particular to
find an explanation for the highest energy events observed, with 
$E\simeq 2$--$3\times 10^{20}$~eV. The result obtained was that 
the reduction by an order of magnitude of the IR background density 
with respect to the original estimates adopted in the study of 
these processes \cite{pu76} implied essentially an increase by an 
order of magnitude in the attenuation length of UHCR nuclei.
It was then concluded that ``the highest energy CR induced 
air-showers could
have been produced by UHCR nuclei propagating from a distance of the order
of 100 Mpc!'', significantly enlarging the volume where possible 
UHCR sources could be located.

We want to point out here that the estimate of this effect is not
appropriate, and that CR nuclei with $E\ga 2\times 10^{20}$~eV cannot 
reach the Earth from distances beyond $\sim 10$~Mpc. Indeed, 
ref.~\cite{st97} just adapted the results of ref.~\cite{pu76} on 
the attenuation of the UHCR spectrum, assigning to the spectrum of CR 
nuclei coming from sources at distances of 100 and 300~Mpc 
(corresponding to travel times of $\sim 10^{16}$ and $3\times 
10^{16}$~s) those previously obtained for travel times of $10^{15}$
 and $3\times 10^{15}$~s. This would have been correct if the dominant
 opacity came from the interactions with the IR photons. However, for 
energies above $\sim 10^{20}$~eV the main opacity source arises from 
photodisintegrations with photons of the  cosmic microwave background 
(CMB), which 
density is well established, rather than with the IR ones. Hence, the 
attenuation changes in a less pronounced way than what was assumed in 
ref.~\cite{st97}.

\vskip -0.2cm
\begin{figure}
\centerline{\hskip 0.3cm\psfig{file=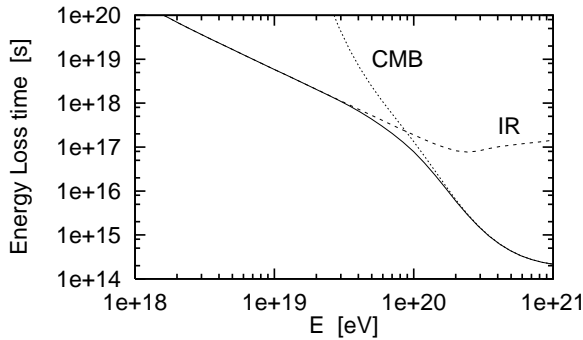,height=1.75in}}
\vskip 0.2cm
\caption[]{Effective energy loss time for photodisintegration of nuclei 
from
microwave photons (CMB), from infrared photons (IR) and 
the total one (solid line).}
\label{Figure 1}
\end{figure}

We show in figure 1 the effective energy loss time for photodisintegration
from microwave photons and from IR ones, 
assuming that the IR background density is a factor 10 smaller than the
`High IR' (HIR) flux of ref.~\cite{pu76}, corresponding to the 
assumption made in ref.~\cite{st97}. It is apparent that the scattering 
from CMB photons cannot be neglected above $E\sim10^{20}$~eV.

In figure 2 we show the attenuation for an $E^{-3}$ input 
spectrum of Fe nuclei, computed as in ref.~\cite{pu76}. 
The dashed lines are for the HIR 
flux of ref.~\cite{pu76} while the 
solid lines are for an IR background density smaller by an order of 
magnitude\footnote{We have not included in these figures the effect of 
energy losses due to pair creation, which should further suppress 
(slightly) the end of the spectra for travel times 
$\sim 3\times 10^{16}$~s. Also not shown are the small tails due 
to secondary nucleons.}.
 Curves from left to right correspond to travel times of
$2\times 10^{17}$, $3\times 10^{16}$, $10^{16}$, $3\times 10^{15}$ and 
$10^{15}$~s. The scaling adopted in ref.~\cite{st97} is not observed,
since for $E>8\times 10^{19}$~eV it is the CMB photons
which provide the main opacity source. For sources at distances of 
10~Mpc, the cutoff energy is already below $2\times 10^{20}$~eV, and it 
is below $10^{20}$~eV for distances $\sim 100$~Mpc.

\vskip -0.1cm
\begin{figure}
\centerline{\hskip 0.3cm\psfig{file=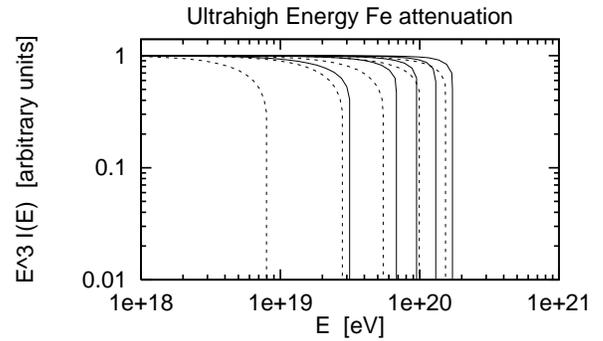,height=1.75in}}
\vskip 0.2cm
\caption[]{Attenuated $E^{-3}$ differential spectrum for ultrahigh 
energy Fe
nuclei for propagation times, from left to right, of $2\times 10^{17}$, 
$3\times 10^{16}$, $10^{16}$, $3\times 10^{15}$ and 
$10^{15}$~s. Dashed lines are for the HIR density, solid lines
for a density 10 times smaller.}
\label{Figure 2}
\end{figure}

Work partially supported by CONICET, Argentina.

\end{document}